\documentclass[11pt,a4paper]{article}
\usepackage[utf8]{inputenc}
\usepackage[export]{adjustbox}

\usepackage{epic,eepic,epsfig,amsmath,amssymb,amsthm}
\usepackage{t1enc}

\usepackage[francais,english]{babel}

\usepackage{caption}
\usepackage{array}

\usepackage{color}

\usepackage{hyperref}

\usepackage{bbm}

\def\virgp{\raise 2pt\hbox{,}}

\renewcommand{\geq}{\geqslant}
\renewcommand{\leq}{\leqslant}
\def\N{{\mathbb N}}

\def\R{{\mathbb R}}

\def\virgp{\raise 2pt\hbox{,}}
\def\cdotpv{\raise 2pt\hbox{;}}

\def\1{\mathbbm{1}}

\newtheorem{theorem}{Theorem}[section]

\newtheorem{proposition}[theorem]{Proposition}
\newtheorem{lemma}[theorem]{Lemma}

\newtheorem{pte}[theorem]{Property}

\theoremstyle{remark}
\newtheorem{remark}{Remark}[section]

\theoremstyle{definition}
\newtheorem{definition}{Definition}[section]

\newtheorem*{notation}{Notation}

\newtheorem*{notations}{Notations}

\theoremstyle{definition}

\theoremstyle{definition}

\newtheorem{assumption}[theorem]{Assumption}

\begin{document}
	
	\title{The inverse Black-Scholes problem in Radon measures space revisited: towards a new measure of market uncertainty}
	
	\author{Nizar Riane$^\dag\,^\ddag$}

	\maketitle
	\centerline{$^\dag$ Universit\'e Mohammed V de Rabat, Maroc\footnote{nizar.riane@gmail.com} }
	\vskip 0.5cm
	
	\centerline{$^\ddag$ Sorbonne Universit\'e}
	
	\centerline{CNRS, UMR 7598, Laboratoire Jacques-Louis Lions, 4, place Jussieu 75005, Paris, France\footnote{Claire.David@Sorbonne-Universite.fr}}
	
	\begin{abstract}
	In this paper, we revisit the inverse Black-Scholes model, the existence of the solution is proved in more rigorous way, and the empirical study is done using different approach based on finite element method.
	
	The article leads to a measure of incertitude in the option market.
	\end{abstract}

	\maketitle
	
	\vskip 1cm
	

	\vskip 1cm
	
	
	\vskip 1cm

	\section{Introduction}

\noindent In 1997, R. Merton and M. Scholes for their works, in particular, the Black-Scholes model \cite{BlackScholes1973}, which continue to be a reference for the option pricing.\\

\noindent Despite the genius behind this construction, the model present many weaknesses when he confront the data, in particular, when the model underprices the option out of the money, and overprices in the money \cite{Franke1999}.\\

\noindent Another weakness of the model is that consider that the market to be isotropic in physicians terminology, while the option price can be affected by incertitude about the future, leading to anisotropic reactions.\\

\noindent In \cite{RDDBS2020}, \cite{RDIBS2021}, we introduced an upgraded version of this model, satisfying the economic foundations of finance and taking into account an important market factor : incertitude.\\

\noindent In this paper, we revisit the inverse Black-Scholes model, as presented in \cite{RDDBS2020}, we give a more rigorous justification of the existence of the solution, and we fit the data using different approach based on finite element method.\\

\noindent At the end of this work, we will prove that the measure based Black-Scholes model is a better candidate to fit financial data, in addition, we will get an interpretation in terms of market incertitude.
	
	\vskip 1cm

	\section{The measure based Black-Scholes formula}

\noindent Next, we will refer to the following functional spaces:

\vskip 1cm 

	\begin{notations}[\textbf{Sobolev spaces}]
		
		Given a strictly positive integer~$d$, a subset~$E$ of~$\R^d$,~$k\, \in\, \N$, and~$p \geq 1$, we recall that the classical Sobolev spaces on~$E$ are:
		
		$$W^{k,p} \left ( E  \right)=
		\left \lbrace f \,\in\, L^p  \left ( E \right) \, , \, 
		\forall\, j \leq k\,: \, D^j f  \,\in\, L^p  \left ( E \right) \right \rbrace   $$
		
		\noindent and:

		$$H^k \left ( E  \right)=W^{k,2} \left ( E  \right)=
		\left \lbrace f \,\in\, L^2  \left ( E \right) \, , \, 
		\forall\, j \leq k\,: \, D^j f  \,\in\, L^2  \left ( E \right) \right \rbrace  \, \cdot$$
		
		\noindent The~ subspace~\mbox{$H_0^k $} of functions which vanish on~$\partial E$ is:
		$$H_0^k \left ( E  \right)=W^{k,2} \left ( E  \right)=
		\left \lbrace f \,\in\, L^2  \left ( E \right) \, , \, f_{|\partial E}=0\,  \text{and: }
		\forall\, j \leq k\,: \, D^j f  \,\in\, L^2  \left ( E \right) \right \rbrace  \, \cdot$$

	\end{notations}

	\vskip 1cm 

\noindent The measure based Black-Scholes model was introduced in \cite{RDIBS2021} in the following variational form:
	
	$$\left \lbrace 
	\begin{array}{ccccc}
	\displaystyle \frac{\partial u}{\partial t}(t,x) \,d\mu&=& \left(- r(t) \,  x \,\displaystyle  \frac{\partial u}{\partial x}(t,x) - \displaystyle \frac{\sigma^2(t,x)}{2}\, x^2\, \Delta u(t,x)  + r(t) \,u(t,x) \right) \, dx &\quad \forall \, t\, \in \,\left[0,T\right], \,\forall \, x \, \in \, \left[0,M\right[ \\
	u(T,x)&=&h(x), &\quad \forall \,x \,\in \, \left[0,M\right[
	\end{array}
	\right.$$
	
	\noindent where~\mbox{$\sigma>0$} represents the volatility,~\mbox{$r>0$} the risk-free interest rate,~\mbox{$T>0$} the maturity of the option and~$u$  the option price. The underlying asset price is supposed to be bounded by some strictly positive number~\mbox{$ M\gg 1$}.\\
	
	\noindent We suppose the measure~$\mu$ to be any finite Radon measure supported in $\overline{\mathcal{M}}=\left[0,M\right]\subset\R^+$ and satisfies the following assumptions
	
	\vskip 1cm 
	
	\begin{assumption}{\ }\\
		\label{ExistenceCondition}
		
		\item There exists two strictly positive constant~\mbox{$C_0$} and~\mbox{$C_1$} such that:

		$$ \forall\, u\, \in \, \mathcal{D}(\mathcal{M}):\quad \parallel u \parallel_{L^2_{\mu}(\mathcal{M})} \leq C_0 \, \left \| x \, \frac{\partial u}{\partial x} \right \|_{L^2(\mathcal{M})}  
		\quad \text{and} \quad  \parallel u \parallel_{L^2(\mathcal{M})} \leq C_1 \, \parallel u \parallel_{L^2_{\mu}(\mathcal{M})}  \, \cdot$$
		
		\noindent where~\mbox{${\cal D}(\mathcal{M})$} denote the space of test functions on~$\mathcal{M}=\left[0,M\right[$, i.e. the space of smooth functions with compact support in~$\mathcal{M}$.
	\end{assumption}
	
	\vskip 1cm
	
\begin{remark}{\ }\\
\noindent We proved In~\cite{RDIBS2021} the first assumption for all finite Radon measures, and we proved the second assumption for all absolute continuous measures with respect to the Lebesgue one.\\
\end{remark}

\vskip 1cm

\noindent We recall the following restrictions on the parameters $r$ and $\sigma$ to ensure existence and uniqueness results:

	\vskip 1cm
	
	\begin{assumption}{\ }\\
		\begin{enumerate}
			\item The risk-free interest rate is bounded on~$\left[0,T\right]$:
			
			\begin{align*}
			\forall\, t\,\in\, [0,T]: \quad 0\leq r\leq r(t)\leq R
			\end{align*} 
			\item There exist two positive constants,~$\underline{\sigma}$ and $\overline{\sigma}$, such that:
			
			\begin{align*}
			\forall\, t\,\in\, [0,T],\, \forall \, x\, \in\, \left[0,M\right]: \quad 0<\underline{\sigma}\leq{\sigma}\leq\overline{\sigma} \, \cdot 
			\end{align*} 
			
			\item There exists a positive constant~$C_\sigma$ such that:
			
			\begin{align*}
			\forall\, t\,\in\, [0,T],\, \forall \, x\, \in\, \left[0,M\right]\, : \quad \left |x \, \frac{\partial \sigma}{\partial x}(t,x)\right |&\leq C_\sigma \, \cdot \\
			\end{align*} 
			
		\end{enumerate}
		
	\end{assumption}

	\vskip 1cm
	
	\noindent The existence and uniqueness of the solution proof can be found in \cite{RDDBS2020}.
	
	\vskip 1cm
	
	\begin{notations}{\ }\\
		
		\noindent Set:
		
		$$ V_{\mathcal{M}}= \left \lbrace  v \, \in \, L^2(\mathcal{M}) \quad ,  \quad x\, \displaystyle \frac{\partial v}{\partial x}\, \in \, L^2(\mathcal{M}) \right\rbrace \quad , \quad 
		W_{\mathcal{M}}= \left \lbrace  v \, \in \, L^2( \mathcal{M}) \quad ,  \quad x^2\, \displaystyle \frac{\partial^2 v}{\partial x^2}\, \in \, L^2(\mathcal{M}) \right\rbrace   $$
		
		\noindent The dual space of~\mbox{$ V_{\mathcal{M}}$} will be denoted by~\mbox{$ V_{\mathcal{M}}^\star$}.
		
	\end{notations}
	
	\vskip 1cm 

	\begin{proposition}[\textbf{Poincar\'e's inequality} ] $\, $ \\
		
		\noindent The space~$\mathcal{D}(\mathcal{M})$ is dense in~$V_{\mathcal{M}}$, and, for any~\mbox{$v\, \in\, \left (\mathcal{D}(\mathcal{M})\right)$},  the following inequality is satisfied:

		$$ \|  v \| _{L^2(\mathcal{M})} \leq 2 \, \left \| x\, \displaystyle \frac{dv}{dx} \right\|_{L^2(\mathcal{M})} $$
		
		\noindent This inequality induces a second norm on $V_{\mathcal{M}}$, given, for any~$v$ in~$V_{\mathcal{M}}$, by:
		
		$$ | v |_{V_{\mathcal{M}}} = \left \|  x\frac{dv}{dx} \right \| _{L^2(\mathcal{M})} \, \cdot$$
		
	\end{proposition}
	
	\vskip 1cm
	
	\begin{proposition}{\textbf{(Continuity and G\r{a}rding inequality) \cite{Achdou2005}} }\\
		
		\noindent The bilinear form~$B(\cdot ,\cdot)$ is continuous on $V_{\mathcal{M}}$, and satisfies the G\r{a}rding inequality:

		$$
		\forall\, u\, \in\, V_{\mathcal{M}}: \quad 
		B(u)  \geq  \displaystyle \frac{\underline{\sigma}^2}{4} | u |_{V_{\mathcal{M}}}^2 - \lambda \, \parallel u \parallel_{L^2(\mathcal{M})}^2   
		$$
		
		\noindent where~$\lambda$ denotes a strictly positive constant.
		
	\end{proposition}
	
	\vskip 1cm

	\begin{remark}{\ }\\
		
		\noindent Given~$u$ in~\mbox{$\text{dom}\, (L)$}, we use the transformation trick~\cite{Zeidler1990}:
		
		$$\tilde{\lambda}=\lambda\,C_1 \quad ,\quad  w=e^{-\tilde{\lambda}\, (T-t)}u$$
		
		\noindent  for the G\r{a}rding constant~$\lambda$ and~$C_1$ at stake in the second conjecture~\ref{ExistenceCondition}. Then:

		$$ \forall \, v\, \in \, \text{dom}\, (L)  :
		\displaystyle \int_{\mathcal{M}} \frac{d}{dt}w(t,x)  \, v(x) \, d\mu  = B(w,v)+\tilde{\lambda} \, \int_{\mathcal{M}}  w(t,x) \,  v(x) \, d\mu =\widehat{B}(w,v) \quad \text{and} \quad 
		w(T,x) =h(x) \, \cdot
		$$
		
		\noindent Without affecting the solution spaces, one obtains the continuity and the coercivity of the form $\widehat{B}$:
		
		\begin{align*}
		|\widehat{B}(u,v)|&\leq 
		C \,  |u|_{V_{\mathcal M}}\, |v|_{V_{\mathcal M}} +\tilde{\lambda}\,  \parallel u \parallel_{L^2_{\mu}}\,  \parallel v \parallel_{L^2_{\mu}}\\
		&\leq(C+\tilde{\lambda}\,  C_0)\,  |u|_{V_{\mathcal M}}\, |v|_{V_{\mathcal M}} \\
		\end{align*} 
		
		\noindent and:
		
		\begin{align*}
		\widehat{B}(u,u)&\geq 
		\displaystyle \frac{\underline{\sigma}^2}{4} \, | u |_{V_{\mathcal M}}^2 - \lambda  \parallel u \parallel_{L^2(\mathcal{M})}^2  + \tilde{\lambda} \, \parallel u \parallel_{L^2_{\mu}}^2 \\
		&\geq  \frac{\underline{\sigma}^2}{4} |  u |_{V_{\mathcal M}}^2\\
		\end{align*} 
		
	\end{remark}
	
	\vskip 1cm
	
	\noindent The following result follows directly from \cite{Zeidler1990}.
	
	\vskip 1cm
	
	\begin{theorem}[\textbf{Measure based Black-Scholes weak solution}] $\, $\\
		
		\noindent Let us define the Gelfand triple (or equipped Hilbert space)~\mbox{$  V_{\mathcal{M}}\subset L^2_{\mu}(\mathcal{M})\subset   V_{\mathcal{M}}^\star$}. For~$h$ in~\mbox{$ V_{\mathcal{M}}$}, the measure based Black-Scholes problem admits a unique weak solution. Moreover, for~\mbox{$k \geq 1$}, the solution map:
		
		\begin{align*}
		L^2_{\mu}(\mathcal{M}) & \rightarrow W^{k,2} \left ([0,T] ; V_{\mathcal{M}} \right ) \\
		h & \mapsto u\\
		\end{align*}
		
		\noindent is continuous.
	\end{theorem}
	
	\vskip 1cm
	
	\begin{theorem}{\textbf{$L^2_\mu$ Regularity} }\\
		
		\noindent We have the estimate, for $0<t<T$:
		
\begin{align*}
\left\| h(x) \right\| _{L^2_\mu(\mathcal{M})}^2  &\geq e^{-2\tilde{\lambda}\, (T-t)} \left\|  u(t,x) \right\|_{L^2_\mu(\mathcal{M})}^2 + \frac{\underline{\sigma}^2}{2}\int_t^T e^{-2\tilde{\lambda}\, (T-s)}  |  u(s,x) |_{V_{\mathcal M}}^2 \,ds\\
\end{align*}
		
	\end{theorem}
	
	\vskip 1cm

\begin{proof}{\ }\\
\noindent Using the remark before by setting $v=e^{-2\tilde{\lambda}\, (T-t)}u$ and integrating the variational formula between $t$ and $T$:

\begin{align*}
\int_t^T \int_{\mathcal{M}}\frac{d}{ds} u(s,x) \, e^{-2\tilde{\lambda}\, (T-s)} \, u(s,x) \, d\mu \, ds &=\int_t^T \frac{d}{ds} \int_{\mathcal{M}} e^{-2\tilde{\lambda}\, (T-s)} \frac{|u(s,x)|^2}{2} \, d\mu \, ds\\
& -\tilde{\lambda} \int_t^T \int_{\mathcal{M}} e^{-2\tilde{\lambda}\, (T-s)} \, |u(s,x)|^2 \, d\mu \, ds\\
&=\frac{1}{2}\parallel u(T,x) \parallel_{L^2_{\mu}(\mathcal{M})}^2 - \frac{e^{-2\tilde{\lambda}\, (T-t)}}{2} \parallel u(t,x) \parallel_{L^2_{\mu}(\mathcal{M})}^2\\
& -\tilde{\lambda} \int_t^T e^{-2\tilde{\lambda}\, (T-s)} \,\parallel u(s,x) \parallel_{L^2_{\mu}(\mathcal{M})}^2 \, ds\\
&=\int_t^T B(u,e^{-2\tilde{\lambda}\, (T-s)}u)\,ds\\
&\geq  \int_t^T \frac{\underline{\sigma}^2}{4} e^{-2\tilde{\lambda}\, (T-s)} | u |_{V_{\mathcal{M}}}^2\,ds - \lambda \, \int_t^T e^{-2\tilde{\lambda}\, (T-s)} \parallel u \parallel_{L^2(\mathcal{M})}^2 \,ds\\
&\geq  \int_t^T \frac{\underline{\sigma}^2}{4} e^{-2\tilde{\lambda}\, (T-s)} | u |_{V_{\mathcal{M}}}^2\,ds - \tilde{\lambda} \, \int_t^T e^{-2\tilde{\lambda}\, (T-s)} \parallel u \parallel_{L^2_{\mu}(\mathcal{M})}^2 \,ds\\
\end{align*}

\noindent The last inequality follows from the second assumption.
\end{proof}

	\vskip 1cm

	\begin{theorem}{\textbf{$V_{\mathcal M}$ Regularity} }\\
		
		\noindent For $h\in V_{\mathcal M}$, we have the estimate:
		
\begin{align*}
 |  u |_{V_{\mathcal M}}^2 &\leq \exp\left(\left(\tilde{\lambda}-\frac{\underline{\sigma}^2}{2C_0}\right)(T-t)\right) | h |_{V_{\mathcal M}}^2 \\
\end{align*}
		
	\end{theorem}
	
	\vskip 1cm
	
	\begin{proof}{\ }\\
\noindent Using the transformation $w=e^{-\tilde{\lambda}\, (T-t)}u$:

\begin{align*}
\frac{\underline{\sigma}^2}{4}| w |_{V_{\mathcal{M}}}^2 \leq \widehat{B} (w)=\frac{1}{2}\frac{\partial}{\partial t} \parallel w \parallel_{L^2_{\mu}(\mathcal{M})}^2 \leq \frac{C_0}{2} \frac{\partial}{\partial t}  | w |_{V_{\mathcal{M}}}^2\\
\end{align*}

\noindent leading to the differential form

\begin{align*}
\begin{cases}
\frac{\partial}{\partial t}  | w |_{V_{\mathcal{M}}}^2 &\geq \frac{\underline{\sigma}^2}{2C_0}| w |_{V_{\mathcal{M}}}^2  \\
| w(T) |_{V_{\mathcal{M}}}^2 &= | h |_{V_{\mathcal{M}}}^2
\end{cases}
\end{align*}

\noindent By Gronwall lemma

\begin{align*}
|  w |_{V_{\mathcal M}}^2 &\leq \exp(-\frac{\underline{\sigma}^2}{2C_0}(T-t))  | h |_{V_{\mathcal M}}^2 \\
\end{align*}

\noindent Back to the original form:

\begin{align*}
 |  u |_{V_{\mathcal M}}^2 &\leq \exp\left(\left(\tilde{\lambda}-\frac{\underline{\sigma}^2}{2C_0}\right)(T-t)\right) | h |_{V_{\mathcal M}}^2 \\
\end{align*}

\end{proof}

	\vskip 1cm
	
	\begin{theorem}{\textbf{Maximum principle} }\\
		
		\noindent If $h\geq0$ then $u\geq 0$ $\mu$-almost everywhere on $\mathcal{M}$.
		
	\end{theorem}
	
	\vskip 1cm

\begin{proof}{\ }\\
\noindent Set $w(t,x)=e^{-\tilde{\lambda}\, (T-t)}u(T-t,x)$. We use Stampacchia truncation \cite{Brezis1983} : Let $G\in C^1(\R)$ such that

\begin{enumerate}
\item $|G'(r)|\leq M$, for some constant $M$.
\item $G$ is increasing on $\left]0,+\infty\right[$.
\item $G(r)=0$ for $r\leq 0$.
\end{enumerate}

\noindent Define, for $s\in \R$, the function

\begin{align*}
H(s)&=\int_0^s G(r) \,dr
\end{align*}

\noindent Set $\phi$ to be

\begin{align*}
\phi(t)=\int_0^M H(-w(t,x)) \, d\mu 
\end{align*}

\noindent Then

\begin{align*}
\phi\in C(\left[0,T\right];\R), \quad \phi(0)=0, \quad \phi\geq 0 \text{ on } \left[0,T\right], \quad \phi\in C^1(\left]0,T\right];\R)
\end{align*}

\noindent and

\begin{align*}
\phi'(t)&=-\int_0^M G(-w(t,x))\frac{d}{dt}w(t,x) \, d\mu \\
&=\widehat{B}(w,G(-w(t,x)))
\end{align*}

\noindent Take $G(r)=r^+$, the positive part, then

\begin{align*}
\phi'(t)&=\widehat{B}(w,G(-w))\\
&=\widehat{B}(w,w^-)\\
&= -\widehat{B}(w^-)\\
&\leq 0
\end{align*}

\noindent Which means that $\phi(t)=0$, then $w \geq 0$ $\mu$-almost everywhere on $\mathcal{M}$.

\end{proof}
	
	\vskip 1cm

	\section{The inverse problem}

\noindent For normalization purpose, we restrict ourselves to finite Radon measure with total mass equals to $M$, this enable ones to take into account the classical Black-Scholes model. For simplicity, we consider probability measures $\mu$ on $\mathcal{M}$, then we multiply $\mu$ by the constant $M$ to get $\mu(\mathcal{M})=M$.

\vskip 1cm

	\begin{definition}[\textbf{Inverse problem}] $\, $\\
		
		We recall the inverse problem as defined in \cite{RDIBS2021}, which consists in finding the measure~$\mu$ associated with the Black-Scholes equation, given a very small parameter~\mbox{$ \varepsilon >0$}, and a noisy measure of the solution~\mbox{$u_{\mu}^{\varepsilon}\in\R^n$} of the solution~\mbox{$u_{\mu}$} such that:
		
		\begin{align*}
		\parallel u_{\mu}^{\varepsilon} - u_{\mu} \parallel &\leq  \varepsilon \, \cdot 
		\end{align*}
	\end{definition}
	
	\vskip 1cm

	\begin{notation}
		
		We denote by~\mbox{$\mathcal{R}(\mathcal{M})$} (respectively $\mathcal{P}(\mathcal{M})$) the space of finite Radon (resp. probability) measures on~\mbox{$\left[0,M\right]$}, and by~\mbox{$C(\mathcal{M})$} the space of (uniformly) continuous functions on~\mbox{$\left[0,M\right]$}.

	\end{notation}

	\vskip 1cm
	
	\begin{proposition}{\textbf{\cite{Bredies2013}} }\\
		
		\noindent The space of finite Radon measures~\mbox{$\mathcal{R}(\mathcal{M})$}  equipped with the \textbf{total variation norm}~\mbox{$\parallel \cdot \parallel_{\mathcal{TV}}$}:
		
		\begin{align*}
		\parallel \mu \parallel_{\mathcal{TV}} = \sup \left\{ \sum_{i=1}^{\infty} |\mu(E_i)| \, ,  \quad
		\underset{  i=1}{\overset{+ \infty}{\bigcup}}\,   E_i =  \left[a,b\right], \quad \{E_i\}\text{ are disjoint and measurable}\right\}
		\end{align*}
		
		\noindent is complete and separable, and is also the dual space of~\mbox{$\mathcal{C}(\mathcal{M})$}:
		
		\begin{align*}
		C(\mathcal{M})^{\star} =\mathcal{R}(\mathcal{M}) \, \cdot\\
		\end{align*}
		
	\end{proposition}
	
	\vskip 1cm
	
	\begin{definition}{\ }\\
		\noindent A sequence of measures~\mbox{$\{\mu_n\}_{n\in \N}\, \in\, \mathcal{R}(\mathcal{M})^\N$}  converges in the weak~sense~$^{\star}$ towards a measure~$\mu$ if and only if
		
		\begin{align*}
		\int_a^b \phi \, d\mu_n \rightarrow \int_a^b \phi \, d\mu \qquad \forall \phi \in C(\mathcal{M})\, \cdot \\
		\end{align*}
		
		\noindent Moreover, every bounded sequence in~\mbox{$\mathcal{R}(\mathcal{M})$} has a weak~$^{\star}$ convergent subsequence.
	\end{definition}
	
	\vskip 1cm
	
	\begin{remark}[\textbf{\cite{Bredies2013}}] $\, $\\
		
		\noindent For~\mbox{$d \geq 1$}, let~\mbox{$\Omega \subset\R^d$} be an open non-empty subset. Then:
		
		\begin{itemize}
			\item[\emph{i}.] the space of all finite linear combinations of $\delta$-peaks 
			
			\begin{align*}
			\left\{ 
			\mu=\sum_{i=1}^n v_i \delta_{x_i} \, ,  \quad n\, \in\, \N , \, v_i \in \R^d, \, x_i \in \Omega, \, i=1,\hdots,n
			\right\}
			\end{align*}
			\item[\emph{ii}.] the space~\mbox{$L^2(\Omega)$}, equipped with the injection~\mbox{$w \rightarrow w\,\mu_{\cal L}$}, for the Lebesgue measure~\mbox{$\mu_{\cal L}$},
		\end{itemize}
		
		\noindent are weakly$^{\star}$ dense subsets of~\mbox{$\mathcal{R}(\mathcal{M})$}.
	\end{remark}
	
	\vskip 1cm	
	
		\begin{proposition}{\ }\\
		
		\noindent The space of probability measures~\mbox{$\mathcal{P}(\mathcal{M})$}  equipped with the \textbf{Prokhorov metric}~\mbox{$d_P$}:
		
		\begin{align*}
		d_P(\mu,\nu) = \inf \left\{ \alpha>0 \, , \, \mu(A)\leq \nu(A_\alpha)+\alpha \quad \text{and} \quad  \nu(A)\leq \mu(A_\alpha)+\alpha \, , \, \forall A\in \mathcal{B}(\mathcal{M}) \right\}
		\end{align*}
		
\noindent where

\begin{align*}
A_\alpha := \{ x \, : \, d(x;A) < \alpha \text{ if } A \neq \emptyset \, ; \, \emptyset_\alpha := \emptyset \, ; \, \forall \alpha > 0 \}
\end{align*}

\noindent and $\mathcal{B}(\mathcal{M})$ is the Borel $\sigma$-algebra on $\left[0,M\right]$, is a compact metric space for the weak$^{\star}$ topology.
		
	\end{proposition}

	\vskip 1cm

	\begin{remark}{\ }\\
		
	\noindent For normalization purpose, we restrict ourselves to the space~\mbox{$\tilde{\mathcal{P}}(\mathcal{M})=\left\{\mu \in \mathcal{R}(\mathcal{M}) \, : \, \mu(\mathcal{M})=M\right\}$}, since the total measure of $\mathcal{M}$ with respect to Lebesgue measure is $M$. Writing $\mu=M\nu$ for $\nu\in\mathcal{P}(\mathcal{M})$, this space inherit the properties of the space of probability measures $\mathcal{P}(\mathcal{M})$.
	\end{remark}
	
	\vskip 1cm

	\begin{notation}
		
		\noindent In the spirit of~\cite{Bredies2013}, we introduce the solution operator of the measure based Black-Scholes problem
		
		$$\begin{array}{ccccc}
		\Psi &: &\tilde{\mathcal{P}}(\mathcal{M}) &\rightarrow &V_{\mathcal{M}} \subset L^2_\mu(\mathcal{M}) \\
		&& \mu & \mapsto &  u_{\mu}(0,\cdot)\\
		\end{array}$$
		
	\end{notation}
	
	\vskip 1cm
	
	\begin{notations}
		
		\noindent Given two probability measures~\mbox{$\mu $} and~\mbox{$\nu$} in~\mbox{$\tilde{\mathcal{P}}(\mathcal{M})$}, we introduce the respective solutions~\mbox{$u_{\mu}$},~\mbox{$u_\nu$},~\mbox{$u_{\mu+\nu}$} of:

		$$\forall\, v \, \in\, {\cal D} ({\cal M}) \, : \quad 
		\displaystyle \frac{d}{dt} \int_0^M u_\mu \, v \, d\mu  = B(u_\mu,v) \quad , \quad \forall\, v \, \in\, {\cal D} ({\cal M}) \, : \quad 
		\frac{d}{dt} \int_0^M u_\nu \, v \, d\nu= B(u_\nu,v)$$

		\noindent and:

		$$\forall\, v \, \in\, {\cal D} ({\cal M}) \, : \quad 
		\displaystyle\frac{d}{dt} \int_0^M u_{\mu+\nu} \, v \, d(\mu+\nu) =\frac{d}{dt} \int_0^M u_{\mu+\nu} \, v \, d\mu+\frac{d}{dt} \int_0^M u_{\mu+\nu} \, v \, d\nu\\
		= B(u_{\mu+\nu},v)\, \cdot$$

	\end{notations}
	
	\vskip 1cm

	\begin{remark}{\ }\\

\begin{enumerate}
	
\item \textbf{Non linearity}: for any~$v$ in~\mbox{${\cal D} \left ({\cal M}\right)$}:
		
		$$\begin{array}{ccc}
		\displaystyle \frac{d}{dt} \left( \int_0^M u_{\mu+\nu} \, v \, d\mu+ \int_0^M u_{\mu+\nu} \, v \, d\nu \right) &=&B(u_{\mu+\nu},v)\\
		\displaystyle \frac{d}{dt} \left( \int_0^M u_\mu\, v \, d\mu +  \int_0^M u_\nu \, v \, d\nu \right) &=&B(u_\mu,v)+B(u_\nu,v)\\
		\end{array}$$
		
		\noindent Then \mbox{$u_{\mu+\nu}= u_\mu+u_\nu$} implies

\begin{align*}
\frac{d}{dt} \left( \int_0^M u_\nu\, v \, d\mu +  \int_0^M u_\mu \, v \, d\nu \right) &=0
\end{align*}		
		
\noindent Choose $v=u$		
		
\begin{align*}
\frac{d}{dt} \frac{1}{2}\left\| u_\nu\right\|^2 _{L^2_{\mu}(\mathcal{M})} = - \frac{d}{dt} \frac{1}{2}\left\| u_\mu\right\|^2 _{L^2_{\nu}(\mathcal{M})}
\end{align*}		

\noindent Integrate between $t$ ant $T$:

\begin{align*}
\left\| h \right\|^2_{L^2_{\mu}(\mathcal{M})} -\left\| u_\nu \right\|^2 _{L^2_{\mu}(\mathcal{M})} + \left\| h \right\|^2 _{L^2_{\nu}(\mathcal{M})} - \left\| u_\mu \right\|^2 _{L^2_{\nu}(\mathcal{M})} &=0 \\
\end{align*}

\noindent Moreover, for $\alpha\in \R$

\begin{align*}
\alpha \frac{d}{dt} \int_0^M u_{\alpha\nu}\, v \, d\mu  &=B(u_{\alpha\nu},v)
\end{align*}		

\noindent Multiply the identity for $u_\mu$ by $\alpha$

\begin{align*}
\alpha \frac{d}{dt} \int_0^M u_\mu\, v \, d\mu  &=\alpha B(u_\mu ,v)
\end{align*}	

\noindent Clearly $u_{\alpha\nu}\neq \alpha u_{\nu}$ and the related operator $\Psi$ is thus not linear.\\

\item \textbf{Weak$^{\star}$-strong Continuity}: for a sequence of probability measures~\mbox{$\left (\mu_n\right)_{n\in\N}\, \in\, {\tilde{\mathcal{P}}} ({\cal M})^\N$} converging towards a given one~$\mu$, it follows from the regularity of the solutions that the sequences $(\left\|  u_{\mu_n} \right\|^2_{L^2_{\mu_n}(\mathcal{M})})$ and $(| u_{\mu_n} |_{V_{\mathcal M}})$ are bounded. Set $w_{\mu}=e^{-\tilde{\lambda}\, (T-t)}u_{\mu}$ and~$v\in  {\cal D} ({\cal M})$:

\begin{align*}
\frac{d}{dt} \left( \int_0^M w_{\mu_n}\, v \, d\mu_n \right) = \widehat{B}(w_{\mu_n},v)\\
\end{align*}

\noindent Subtracting the inequalities defining $w_{\mu_n}$ and $w_{\mu_m}$ for $m, n\in\N$, and set $v=w_{\mu_n}-w_{\mu_m}$

\begin{align*}
\widehat{B}( w_{\mu_n}-w_{\mu_m})&= \int_0^M \frac{d}{dt} w_{\mu_n}(w_{\mu_n}-w_{\mu_m}) \, d\mu_n - \int_0^M  \frac{d}{dt} w_{\mu_m}(w_{\mu_n}-w_{\mu_m}) \, d\mu_m \\
&= \int_0^M \frac{d}{dt} w_{\mu_n}(w_{\mu_n}-w_{\mu_m}) \, d(\mu_n-\mu_m) + \int_0^M  \frac{d}{dt} (w_{\mu_n}-w_{\mu_m})(w_{\mu_n}-w_{\mu_m}) \, d\mu_m \\
\end{align*}

\noindent Then

\begin{align*}
 \int_t^T \widehat{B}( w_{\mu_n}-w_{\mu_m})\, ds&= \int_t^T \int_0^M \frac{d}{dt} w_{\mu_n}(w_{\mu_n}-w_{\mu_m}) \, d(\mu_n-\mu_m) \, ds + \frac{1}{2} \int_t^T \frac{d}{dt} \left\|  w_{\mu_n}-w_{\mu_m} \right\|_{L^2_{\mu_n}(\mathcal{M})}^2 \, ds \\
&= \int_t^T \int_0^M \frac{d}{dt} w_{\mu_n}(w_{\mu_n}-w_{\mu_m}) \, d(\mu_n-\mu_m) \, ds - \frac{1}{2}  \left\|  w_{\mu_n}-w_{\mu_m} \right\|_{L^2_{\mu_n}(\mathcal{M})}^2 \\
\end{align*}

\noindent Equivalently

\begin{align*}
\frac{1}{2}  \left\|  w_{\mu_n}-w_{\mu_m} \right\|_{L^2_{\mu_n}(\mathcal{M})}^2 + \int_t^T \frac{\underline{\sigma}^2}{4} |  w_{\mu_n}-w_{\mu_m} |_{V_{\mathcal M}}^2\, ds &\leq \frac{1}{2}  \left\|  w_{\mu_n}-w_{\mu_m} \right\|_{L^2_{\mu_n}(\mathcal{M})}^2 + \int_t^T \widehat{B}( w_{\mu_n}-w_{\mu_m})\, ds\\
&= \int_t^T \int_0^M \frac{d}{dt} w_{\mu_n}(w_{\mu_n}-w_{\mu_m}) \, d(\mu_n-\mu_m) \, ds  \\
\end{align*}

\noindent Since $( u_{\mu_n})$ is bounded in $L^2_{\mu_n}(\mathcal{M})$ and $V_\mathcal{M}$, then it is bounded in $L^2_{\mu_m}(\mathcal{M})$, for all $m\in \N$, it follows from Cauchy-Schwartz inequality and the definition of $w_{\mu_n}$ that $( u_{\mu_n})$ is a Cauchy sequence in $V_{\mathcal M}$ and then converges to a limit $u_\mu$.

\end{enumerate}			
	\end{remark}
	
	\vskip 1cm
	
	\begin{definition}[\textbf{Tikhonov regularized solution}] $\, $\\

		\noindent The Tikhonov minimization problem is the solution of
		
		\begin{align*}
		\min_{\mu\in \tilde{\mathcal{P}}(\mathcal{M})} \mathcal{T}_{\alpha}(\mu) =\displaystyle \frac{\parallel  \Psi (\mu) -  u_{\mu}^{\varepsilon}  \parallel^2_{L^2(\mathcal{M})}}{2} + \alpha \parallel \mu \parallel_{\mathcal{TV}}
		\end{align*}
		
		\noindent where~$\alpha$ denotes the regularization parameter.
	\end{definition}
	
	\vskip 1cm

\begin{theorem}{\ }\\
\noindent Thikhonov regularized problem admit a solution.
\end{theorem}

	\vskip 1cm

\begin{proof}{\ }\\
\noindent The operator $\Psi$ is weak$^{\star}$-strong continuous, so is Tikhonov functional from composition properties. Since the space $\tilde{\mathcal{P}}(\mathcal{M})$ is compact for the weak$^{\star}$ topology, the result follows.\\
\end{proof}

\vskip 1cm

\section{Finite element}

\noindent In \cite{RDIBS2021}, we solved the inverse measure problem using the finite difference technique, based on our articles \cite{RianeDavidFDMS}, \cite{RDSGvsAC} and \cite{RianeDavidFVMS}.\\

\noindent In order to construct a finite element approximation, we need to recall the weak form of measure based Black-Scholes equation:

\begin{align*}
 \frac{d}{dt}\,  \int_0^M u \, v \, d\mu&= B(u,v)
\end{align*}
		
\noindent   where~\mbox{$B(\cdot, \cdot)$} is the non-symmetric bilinear form defined, for any pair~\mbox{$(u,v)$} of~\mbox{$V_{\mathcal{M}}\times V_{\mathcal{M}}$}, through:
		
		\begin{align*}
		B(u,v)&= \displaystyle \int_0^M \frac{\sigma^2 x^2}{2} \, \frac{\partial u}{\partial x} \frac{\partial v}{\partial x} dx + \int_0^M \left(\sigma^2+x\,\sigma\, \frac{\partial \sigma}{\partial x}-r\right) x \, \frac{\partial u}{\partial x}\, v \,dx +\int_0^M r\,u\,v\, dx 
		\, \cdot
		\end{align*}

\noindent We will proceed as in \cite{Mautner2007NumericalTO} :

	\vskip 1cm	

\subsection{$\mathbb{P}_1$ elements}
	
	\begin{notation} 
		
		\noindent For~\mbox{$m\geq 1$}, we introduce the uniform subdivision of the set~\mbox{$\overline{\mathcal{M}}=\left[0,M\right]$} into~\mbox{$m$} equidistant points~\mbox{$x_j$},~\mbox{$ j \, \in\, \left \lbrace 0, \hdots, m \right \rbrace$}: 
		
		$$x_j=\displaystyle\left(j\,\frac{M}{m}\right)=j\,\Delta x  \, \cdot$$
		
\noindent for $\Delta x=\displaystyle\frac{M}{m}$. 

	\end{notation} 
	
	\vskip 1cm	
	
\begin{definition}[\textbf{Finite element basis functions} \cite{AllaireLivre2007}] $\, $\\
\noindent Define the vector space

\begin{align*}
V_{\mathcal{M}}^h&=\text{span}\{\phi_1,\hdots,\phi_{m-1}\}\\
\end{align*}

\noindent the space of piecewise linear finite element basis functions where $(\phi_j)_{0\leq j \leq m}$ are given by

\begin{align*}
\phi_0(x)&=\begin{cases}
1,  \quad & x <x_{0}\\
\displaystyle\frac{x-x_{1}}{x_0-x_{1}},  \quad &x_{0}\leq x\leq x_{1}\\
0,   \quad &\text{otherwise}\\
\end{cases}\\
\phi_j(x)&=\begin{cases}
\displaystyle\frac{x-x_{j-1}}{x_j-x_{j-1}},  \quad &x_{j-1}\leq x\leq x_{j}\\
\displaystyle\frac{x-x_{j+1}}{x_j-x_{j+1}},  \quad &x_{j}\leq x\leq x_{j+1}\\
0,   \quad &\text{otherwise}\\
\end{cases}\\
\phi_m(x)&=\begin{cases}
\displaystyle\frac{x-x_{m-1}}{x_m-x_{m-1}},  \quad &x_{m-1}\leq x\leq x_{m}\\
\displaystyle\frac{x-K e^{-rt}}{M-K e^{-rt}},  \quad &x_{m}<x\\
0,   \quad &\text{otherwise}\\
\end{cases}
\end{align*}

\end{definition} 
	
	\vskip 1cm		

\noindent Using the same approach to establish the existence and uniqueness of the solution in $V_{\mathcal{M}}$, we can establish the following result:
	
	\vskip 1cm
	
	\begin{pte}[\textbf{Finite element approximation}] $\, $\\

\noindent The variational problem

\begin{align*}
 \frac{d}{dt}\,  \int_0^M u \, v \, d\mu&= B(u,v)
\end{align*}

\noindent has a unique solution in 	$V_{\mathcal{M}}^h$.
		
	\end{pte}
	
	\vskip 1cm
	
\noindent Write the approximated solution as

\begin{align*}
u^m(t,x)&=\sum_{j=0}^m \beta_j(t) \, \phi_j(x)\\
&=\overline{u^m}(t,x)+\beta_0(t)\,\phi_0(x) + \beta_m(t) \, \phi_m(x)
\end{align*}
	
\noindent for $\overline{u^m}\in V_{\mathcal{M}}^h$. The variational formula can be transformed to

\begin{align*}
\sum_{j=0}^m \beta_j'(t) \int_0^M \phi_j(x) \, \phi_i(x) \, d\mu&= B\left(\phi_j(x),\phi_i(x)\right)\\
&=\sum_{j=0}^m \beta_j(t) \left(\int_0^M \frac{\sigma^2 x^2}{2} \, \phi_j'(x) \, \phi_i'(x) dx + \int_0^M \left(\sigma^2+x\,\sigma\, \frac{\partial \sigma}{\partial x}-r\right) x \, \phi_j'(x) \, \phi_i(x) \,dx \right.\\
&\left. +\int_0^M r\,\phi_j(x) \, \phi_i(x)\, dx\right)\\
\end{align*}

\noindent Set, for $1\leq i, j \leq m-1$:

\begin{align*}
\mathbf{B}^m_j(t)&=\beta_j(t)\\
\mathbf{M}^m_{i,j}&= \int_0^M \phi_j(x) \, \phi_i(x) \, d\mu\\
\mathbf{K}^m_{i,j}&= B\left(\phi_j(x),\phi_i(x)\right)\\
\end{align*}

\noindent More precisely

\begin{align*}
\mathbf{M}^m_{j,j-1}&=\frac{1}{(\Delta x) ^2}\int_{x_{j-1}}^{x_{j}} (x-x_{j})(x_{j-1}-x) \, d\mu\\
\mathbf{M}^m_{j,j}&=\frac{1}{(\Delta x) ^2}\left( \int_{x_{j-1}}^{x_{j}} (x-x_{j-1})^2 \, d\mu + \int_{x_{j}}^{x_{j+1}} (x-x_{j+1})^2 \, d\mu \right)\\
\mathbf{M}^m_{j,j+1}&=\frac{1}{(\Delta x) ^2} \int_{x_{j}}^{x_{j+1}} (x-x_{j+1})(x_{j}-x) \, d\mu\\
\end{align*}

\noindent and

\begin{align*}
\mathbf{K}^m_{j,j-1}&=\frac{1}{(\Delta x) ^2}\left(- \int_{x_{j-1}}^{x_{j}} \frac{\sigma^2 x^2}{2} \, dx - \int_{x_{j-1}}^{x_{j}} \left(\sigma^2+x\,\sigma\, \frac{\partial \sigma}{\partial x}-r\right) x \left(x-x_{j}\right) \,dx-r\int_{x_{j-1}}^{x_{j}} (x-x_{j})(x-x_{j-1}) \, dx\right)\\
\mathbf{K}^m_{j,j}&=\frac{1}{(\Delta x) ^2}\left(\int_{x_{j-1}}^{x_{j+1}} \frac{\sigma^2 x^2}{2} \, dx + \int_{x_{j-1}}^{x_{j}} \left(\sigma^2+x\,\sigma\, \frac{\partial \sigma}{\partial x}-r\right) x \left(x-x_{j-1}\right) \,dx  \right.\\
&\left.+ \int_{x_{j}}^{x_{j+1}} \left(\sigma^2+x\,\sigma\, \frac{\partial \sigma}{\partial x}-r\right) x \left(x-x_{j+1}\right) \,dx +r\int_{x_{j-1}}^{x_{j}} (x-x_{j-1})^2 \,dx + r\int_{x_{j}}^{x_{j+1}}(x_{j+1}-x)^2 \, dx\right)\\
\mathbf{K}^m_{j,j+1}&=\frac{1}{(\Delta x) ^2}\left(- \int_{x_{j}}^{x_{j+1}} \frac{\sigma^2 x^2}{2} \, dx - \int_{x_{j}}^{x_{j+1}} \left(\sigma^2+x\,\sigma\, \frac{\partial \sigma}{\partial x}-r\right) x \left(x-x_{j}\right) \,dx-r\int_{x_{j}}^{x_{j+1}} (x-x_{j})(x-x_{j+1}) \, dx\right)\\
\end{align*}

\noindent We impose the limit conditions $\beta_0(t)=0$ and $\beta_m(t)=M-K e^{-r(T-t)}$. The problem satisfies then the dynamic

\begin{align*}
\begin{cases}
\displaystyle \frac{d}{dt}\mathbf{B}^m \mathbf{M}^m &= \mathbf{B}^m \mathbf{K}^m +\beta_m(t)\mathbf{C}^m\\
\mathbf{B}^m(T)&=\mathbf{H}
\end{cases}
\end{align*}

\noindent where, for $1\leq i,j \leq m-1$,

\begin{align*}
\mathbf{H}_j&=h(x_j)\\
\mathbf{C}_{m}&=\frac{1}{(\Delta x) ^2}\left(- \int_{x_{m-1}}^{x_{m}} \frac{\sigma^2 x^2}{2} \, dx - \int_{x_{m-1}}^{x_{m}} \left(\sigma^2+x\,\sigma\, \frac{\partial \sigma}{\partial x}-r\right) x \left(x-x_{m}\right) \,dx-r\int_{x_{m-1}}^{x_{m}} (x-x_{m})(x-x_{m-1}) \, dx\right)\\
\mathbf{C}_{i\neq m}&=0\\
\end{align*}

\vskip 1cm

\subsection{Convergence and error estimate}

\noindent Let proceed without affecting the solution spaces, as in the remark of the first section by setting $\tilde{\lambda}=\lambda\,C_1$ and $u=e^{\tilde{\lambda}\, (T-t)} w$, where $w$ is the solution of the transformed problem

\begin{align*}
 \frac{d}{dt}\,  \int_0^M w \, v \, d\mu&= \widehat{B}(w,v)
\end{align*}

\noindent The bilinear form $\widehat{B}(w,v)=B(w,v)+\tilde{\lambda} \, \int_{\mathcal{M}}  w(t,x) \,  v(x) \, d\mu$, satisfies continuity and coercivity:
		
		\begin{align*}
		|\widehat{B}(u,v)|&\leq(C+\tilde{\lambda}\,  C_0)\,  |u|_{V_{\mathcal M}}\, |v|_{V_{\mathcal M}} \\
		\widehat{B}(u,u)&\geq  \frac{\underline{\sigma}^2}{4} |  u |_{V_{\mathcal M}}^2\\
		\end{align*} 

\noindent J. L. Lions and E. Magenes theory \cite{Lions1968} implies the existence and uniqueness of the solution for the approximated problem.\\

\noindent We use the decomposition $u^m(t,x)=\overline{u^m}(t,x)+\beta_0(t)\,\phi_0(x) + \beta_m(t) \, \phi_m(x)$ to consider the homogeneous part $\overline{u^m}(t,x)$.\\

\noindent Define the projector $\pi \, : \, V_{\mathcal{M}} \rightarrow V_{\mathcal{M}}^h$ by

\begin{align*}
\pi(v)(x)=\sum_{j=1}^{m-1} v(x_j) \, \phi_j(x)
\end{align*}

\vskip 1cm

\begin{lemma}{\ }\\
\noindent There is a constant $C_2$ independent of $\Delta x$ such that, for all $v\in V_{\mathcal{M}}^h$

\begin{align*}
| v-\pi(v) |_{V_{\mathcal{M}}} &\leq C_2 \, (\Delta x)^2 \left\| (xv(x))'' \right\|_{L^2(\mathcal{M})}
\end{align*}

\end{lemma}

\vskip 1cm

\begin{proof}{\ }\\
\noindent Let $v\in \mathcal{D}(\mathcal{M})$. For $x\in\left]x_j,x_{j+1}\right[$:

\begin{align*}
xv(x)-\pi(xv(x)) &=xv(x)-\left(x_jv(x_j)+\frac{x_{j+1}v(x_{j+1})-x_{j}v(x_j)}{x_{j+1}-x_j}\right)(x-x_j)\\
&=\int_{x_j}^x  \frac{d}{ds} (sv(s)) \,ds-\frac{x-x_{j}}{x_{j+1}-x_j}\left(\int_{x_j}^{x_{j+1}}  \frac{d}{ds} (sv(s)) \,ds\right)\\
&=(x-{x_j}) \frac{d}{ds}(sv(s))(x_j+\alpha) -(x-x_{j})\frac{d}{ds}(sv(s))(x_j+\beta)\\
&=(x-{x_j}) \int_{x_j+\alpha}^{x_j+\beta} \frac{d^2}{ds^2}(sv(s))\,ds\\
\end{align*}

\noindent for $0\leq \alpha \leq x-x_j$ and $0\leq \beta \leq \Delta x$. We get from Cauchy-Schwartz inequality

\begin{align*}
|xv(x)-\pi(xv(x))|^2 &\leq (\Delta x)^2 \left(\int_{x_j}^{x_{j+1}} |\frac{d^2}{ds^2}(sv(s))|\,ds\right)^2\leq (\Delta x)^3 \left(\int_{x_j}^{x_{j+1}} |\frac{d^2}{ds^2}(sv(s))|^2\,ds\right)\\
\end{align*}

\noindent integrating over $\left[x_j,x_{j+1}\right]$ and summing 

\begin{align*}
| v-\pi(v) |^2_{V_{\mathcal{M}}}&=\sum_{j=1}^{m-1}  \int_{x_j}^{x_{j+1}} |xv(x)-\pi(xv(x))|^2 dx\\
&\leq (\Delta x)^4 \sum_{j=1}^{m-1}  \left(\int_{x_j}^{x_{j+1}} |\frac{d^2}{ds^2}(sv(s))|^2\,ds\right)\\
&= (\Delta x)^4 \left\| (sv(s))'' \right\|^2_{L^2(\mathcal{M})}
\end{align*}

\noindent The result follows then by density.

\end{proof}
	
\vskip 1cm

\noindent It follows the following theorem

\vskip 1cm

\begin{theorem}{\ }\\
\noindent Let $u\in V_{\mathcal{M}}$ and $u^m\in V_{\mathcal{M}}^h$ be respectively, the solution and the finite element approximation of the measure based Black-Scholes equation. Then the finite element method converges.\\
\noindent In addition, if $h\in W_{\mathcal{M}}$, then there exist a constant $C(T)$, independent of $\Delta x$, such that

\begin{align*}
\left\| u-u^m \right\|_{L^2_\mu(\mathcal{M})} &\leq C(T) \, (\Delta x)^2 \left\| (xh(x))'' \right\|^2_{L^2(\mathcal{M})}
\end{align*}

\end{theorem}

\vskip 1cm

\begin{proof}{\ }\\
\noindent It suffice to resonate in terms of the transformed problem.\\

\noindent Consider the variational inequalities involving $u\in W_{\mathcal{M}}$ and $u^m\in V_{\mathcal{M}}^h$, by taking $v\in V_{\mathcal{M}}^h$

\begin{align*}
\frac{d}{dt}\,  \int_0^M u \, v \, d\mu&= \widehat{B}(u,v)\\
\frac{d}{dt}\,  \int_0^M u^m \, v \, d\mu&= \widehat{B}(u^m,v)
\end{align*}

\noindent subtract the two inequalities and define the error $\mathfrak{e}=u-u^m$, this yield to the the formula

\begin{align*}
\frac{d}{dt}\,  \int_0^M \mathfrak{e} \, v \, d\mu&= \widehat{B}(\mathfrak{e},v)\\
\end{align*}

\noindent It follows from coercivity and continuity of $\widehat{B}$ that

\begin{align*}
\left\| u-u^m \right\|_{V_{\mathcal{M}}}&\leq C_3 \, \inf_{v^m\in V_{\mathcal{M}}^h } \left\| u-v^m \right\|_{V_{\mathcal{M}}}
\end{align*}

\noindent for some constant $C_3$. Using the regularity solution estimate, the first assumptions and the lemma before, we get for $0\leq t \leq T$

\begin{align*}
\left\|  \mathfrak{e}(t,x) \right\|_{L^2_\mu(\mathcal{M})}  &\leq  e^{\tilde{\lambda}\, (T-t)} \, \left\| \mathfrak{e}(T,x)\right\| _{L^2_\mu(\mathcal{M})}  \\
&\leq  C_0 \, e^{\tilde{\lambda}\, T} \, | \mathfrak{e}(T,x)| _{V_{\mathcal{M}}}  \\
&\leq  C_0 \, C_3 \,e^{\tilde{\lambda}\, T} \, | u(T,x)-\pi(u)(T,x)| _{V_{\mathcal{M}}}  \\
&\leq  C_0 \, C_2 \, C_3\, e^{\tilde{\lambda}\, T} \, (\Delta x)^2 \left\| (xu(T,x))'' \right\|_{L^2(\mathcal{M})}  \\
&\leq  C(T) \, (\Delta x)^2 \left\| (xh(x))'' \right\|_{L^2(\mathcal{M})}  \\
\end{align*}

\end{proof}

\vskip 1cm

\noindent We deduce from the existence and uniqueness of the transformed solution, the existence and uniqueness of the original one, and then it's finite element approximation.

\vskip 1cm

	\subsection{Discretization of the Tikhonov problem}
	
	\hskip 0.5cm Given~\mbox{$N\geq1$},~\mbox{$m\geq1$},~\mbox{$0\leq n \leq N$} and~\mbox{$j\, \in\, \{0,\hdots,m\}$}, we denote by~\mbox{$u^{m}_\mu(t,j)=\sum_{i=0}^m \beta_i(t) \, \phi_i(x_j)$} the finite element approximation of the solution~\mbox{$u_\mu(t,x_j)$}. The coefficient vector $\mathbf{B}^m(t)=(\beta_i(t))_{0\leq i \leq m}$ follows the differential system

\begin{align*}
\begin{cases}
\displaystyle \frac{d}{dt}\mathbf{B}^m \mathbf{M}^m&= \mathbf{B}^m \mathbf{K}^m +\beta_m(t)\mathbf{C}^m\\
\mathbf{B}^m(T)&=\mathbf{H}
\end{cases}
\end{align*}
	
\noindent as stated before. Define the approximation measure as a Dirac linear combination
	
	$$\nu=\displaystyle \sum_{i=0}^{m} \gamma_i \, \delta_i$$
	
\noindent for $\sum_{i=0}^m \gamma_i=M$, which is dense in $\tilde{\mathcal{P}}(\mathcal{M})$. The matrix $\mathbf{M}^m$ simplifies to a diagonal matrix

\begin{align*}
\mathbf{M}^m_{j,j}&=\gamma_j\\
\mathbf{M}^m_{\underset{i\neq j}{j,i}}&=0\\
\end{align*}

\noindent Now consider an empirical measure $u_{\mu}^{\varepsilon}\in\R^n$ of the solution, we introduce the finite element approximation operator $\tilde{\Psi} \, : \, \tilde{\mathcal{P}}(\mathcal{M}) \rightarrow \R^n$, such that:

		$$\tilde{\Psi}(\mu)_i= \int_{\mathcal{M}} \delta_{x_i} u^m_\mu \, dx $$
		
\noindent This construction justify the discretized Tikhonov problem:	

\vskip 1cm
	
	\begin{definition}[\textbf{Discretized Tikhonov problem}] $\, $\\
		
		\noindent We define the \textbf{discretized Tykhonov problem} as
		
		\begin{align*}
		\min_{\gamma\in \R^m, \, \sum \gamma_i =M} \mathcal{J}(\mu)= \frac{\parallel  \tilde{\Psi} (\mu) -  u_{\mu}^{\varepsilon}  \parallel^2}{2} + \alpha \sum_{i=0}^{m} |\gamma_i|
		\end{align*}
		
		\noindent where $\mu=\sum_{i=1}^m \gamma_i \, \delta_i$.
		
	\end{definition}
	
	\vskip 1cm
	
	\begin{theorem}{\ }\\
		\noindent The discretized Tikhonov minimization problem admits a unique solution $\mu\in \tilde{\mathcal{P}}(\mathcal{M})$.
	\end{theorem}
	
	\vskip 1cm
	
	\begin{proof}{\ }\\
		\noindent The existence follows from the continuity of $\tilde{\Psi}$, which follows from the continuous dependence of the solution of the linear differential system
		
\begin{align*}
\begin{cases}
\displaystyle  \frac{d}{dt}\mathbf{B}^m \mathbf{M}^m &=\left(
\mathbf{K}^m \mathbf{B}^m+\beta_m\mathbf{C}^m\right)\\
\mathbf{B}^m(T)&=\mathbf{H}
\end{cases}
\end{align*}
		
\noindent to the matrix $\mathbf{M}^m$ (continuity of the exponential), combined with the form of $u^m$
		
	\begin{align*}
u^m(t,x)&=\mathbf{B}^m(t) \cdot \mathbf{\Phi}^m(x) +\beta_m(t)\phi_m(x)
\end{align*}

\noindent where $\mathbf{\Phi}^m=(\phi_i)_{1\leq i\leq m-1}$, is the finite element basis. The uniqueness follows by strict convexity of Tikhonov functional and the exponential.
	\end{proof}
	
	\vskip 1cm

	\subsection{Empirical results}

\noindent To confront our theory to reality, we will establish comparison the classical and the measure based Black-Scholes models, based on their adjustment qualities of financial data. For this purpose, we used a sub-sample of data from Vance L. Martin~\cite{Martin2005}. The sample consist of~\mbox{$\mathbf{N}=269$} observations on the European call options written on the~\mbox{S$\&$P$500$} stock index on the~\mbox{$4^{\textrm{th}}$} of April, $1995$, we refer to \cite{RDIBS2021} for detailed description.\\
	
\noindent We recall the main characteristics of these options:
	
	\begin{itemize}
		\item[\emph{ii}.]  The strike : $K=520$.
		\item[\emph{iii}.] The maturity : $T=0.457534$.
		\item[\emph{v}.]  The interest rate : $r=0.0591$.
		\item[\emph{vi}.] The volatility : $\sigma=0.076675$.
		
	\end{itemize}
	
	\noindent The table~\ref{TableOption} gives the data range for the option and the stock index prices:
	
	\vskip 1cm
	
	\begin{center}
		
		\begin{table}
			\hskip 5cm
			\begin{tabular}{ |c|c|c| } 
				\hline
				\textbf{Statistic} & \textbf{Option price} & \textbf{Stock price} \\
				\hline
				Max & $0.106250$ & $498.5040$ \\
				Min & $9.50000$ & $496.4980$ \\ 
				Mean & $9.971304$ & $497.3208$ \\ 
				\hline
			\end{tabular}
			\caption{Option and the stock index prices.}
			\label{TableOption}
		\end{table}
	\end{center}

	\vskip 1cm	

\noindent Next, we represent the classical solution versus the measure based one. The approximation parameter is fixed to $m=30$:

	\vskip 1cm
	
	\begin{figure}[!htb]
		\begin{center}
			\includegraphics[scale=1.25]{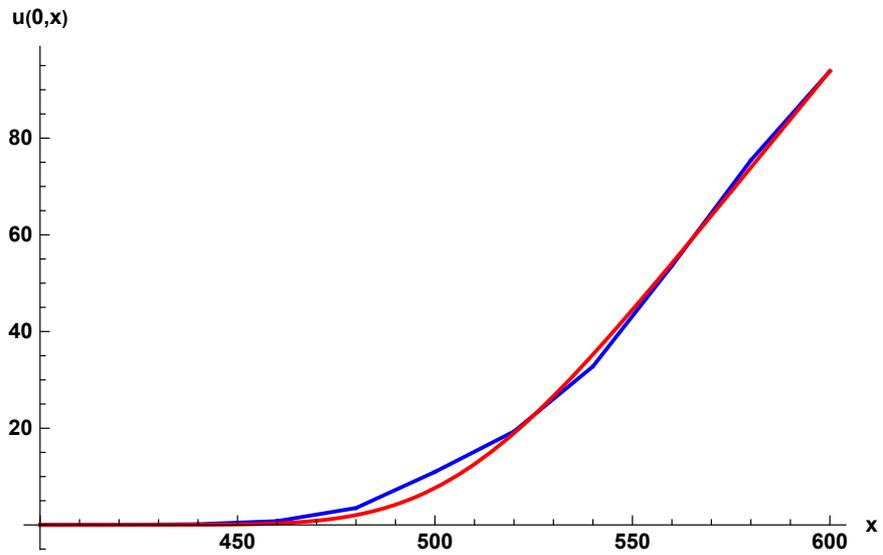}			
			\captionof{figure}{The graph of the measure based Black-Scholes solution (blue) versus the classical Black-Scholes solution (red).}
			\label{GraphSolution}
		\end{center}
	\end{figure}
	
	\vskip 1cm

	\begin{figure}[!htb]
		\begin{center}
		\includegraphics[scale=1.25]{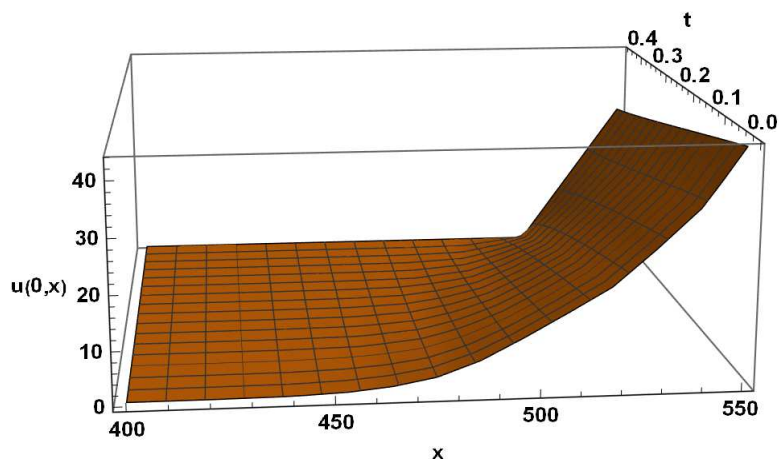}
			\captionof{figure}{3D representation of the measure based Black-Scholes solution.}
			\label{GraphSolution}
		\end{center}
	\end{figure}
	
	\vskip 1cm

\noindent To evaluate the adjustment quality of the measure based model, we plot the two models versus the scatter plot of the data:
	
\vskip 1cm

	\begin{figure}[!htb]
		\begin{minipage}[b]{0.5\linewidth}
			\centering
			\includegraphics[width=\linewidth]{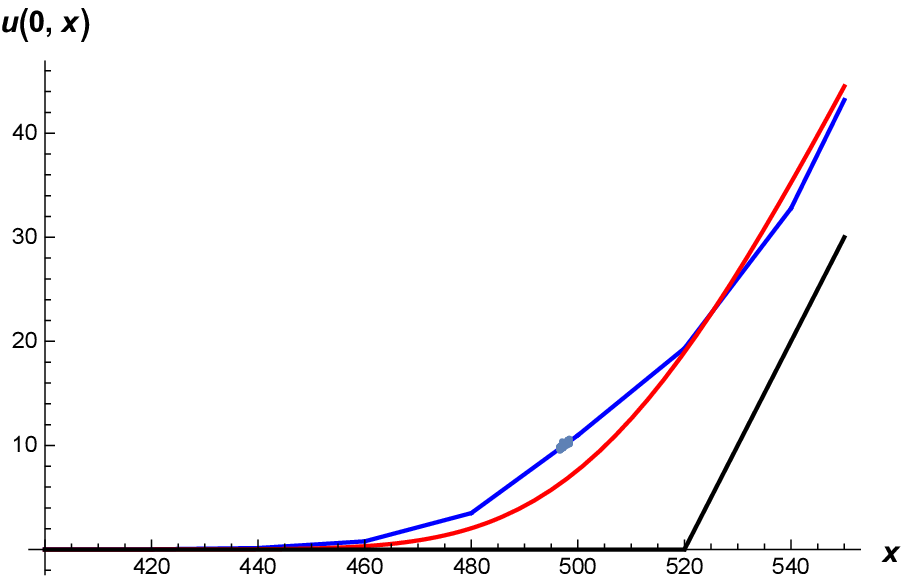} 
			\vspace{4ex}
		\end{minipage}
		\begin{minipage}[b]{0.5\linewidth}
			\centering
			\includegraphics[width=\linewidth]{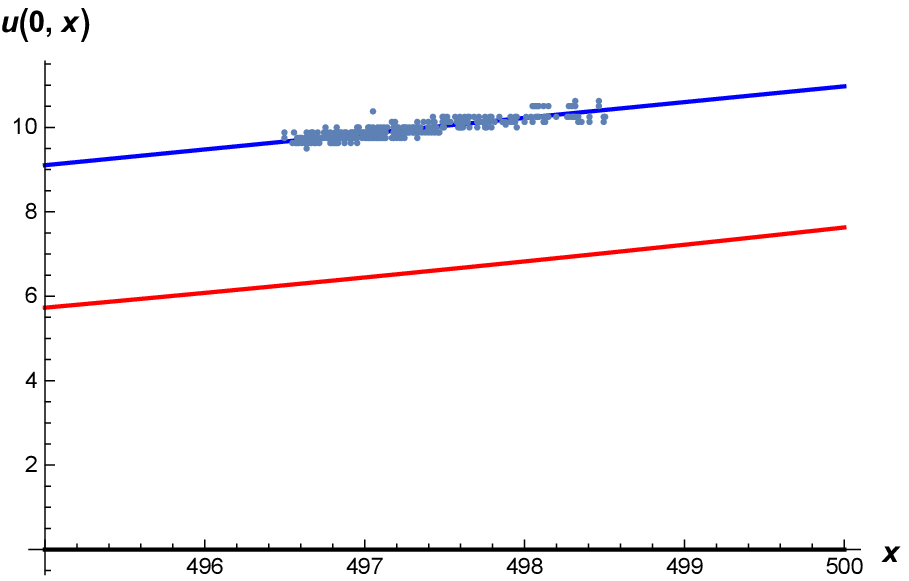}
		\end{minipage}
		\vspace{4ex}
    	\begin{minipage}[b]{0.1\linewidth}
    	\centering
   		\includegraphics[scale=1]{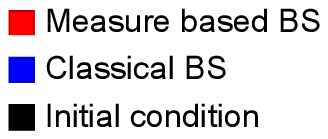} 
    	\vspace{4ex}
  \end{minipage}
		\captionof{figure}{Fitted solution at different scales.}
		\label{Zooms}\end{figure}
		
\vskip 1cm

\noindent The measure based model offer a much better approximation of the solution, which can be measured by Tikhonov functional:

	\vskip 1cm
	
	\begin{center}
		
		\begin{table}[!htb]
			\hskip 5cm
			\begin{tabular}{ |c|c|c| } 
				\hline
				 & \textbf{Classical BS} & \textbf{Measure based BS} \\
				\hline
				\textbf{Tikhonov functional} & $3375.05$ & $4.49$ \\
				\hline
			\end{tabular}
			\caption{Fit quality comparison.}
			\label{TableOption}
		\end{table}
	\end{center}
	
	\vskip 1cm
	
\noindent We represent the parameters $\gamma_i/M$ in terms of a probability density function. To focus on those belonging to the data region, we will use an adaptive mesh:
	
\vskip 1cm

	\begin{figure}[!htb]
		\begin{minipage}[b]{0.5\linewidth}
			\centering
			\includegraphics[width=\linewidth]{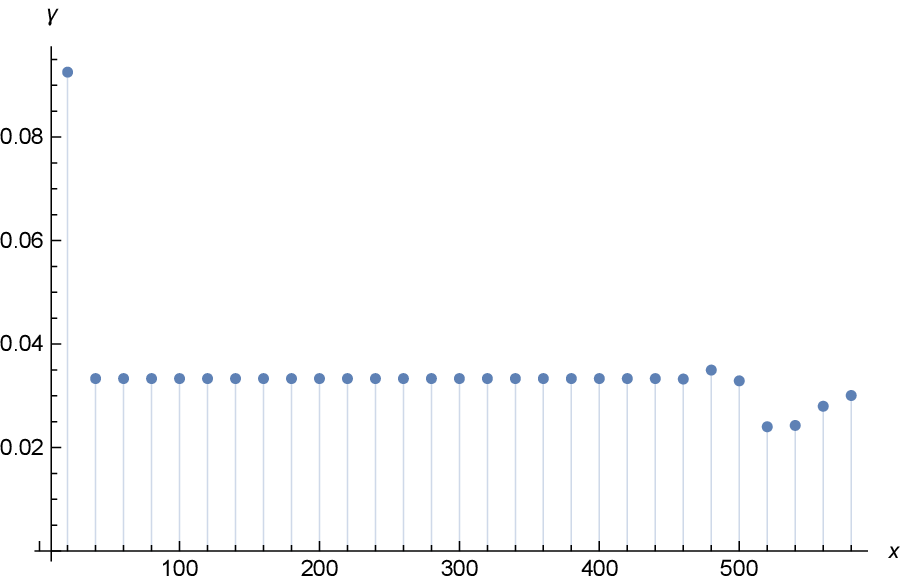} 
			\vspace{4ex}
		\end{minipage}
		\begin{minipage}[b]{0.5\linewidth}
			\centering
			\includegraphics[width=\linewidth]{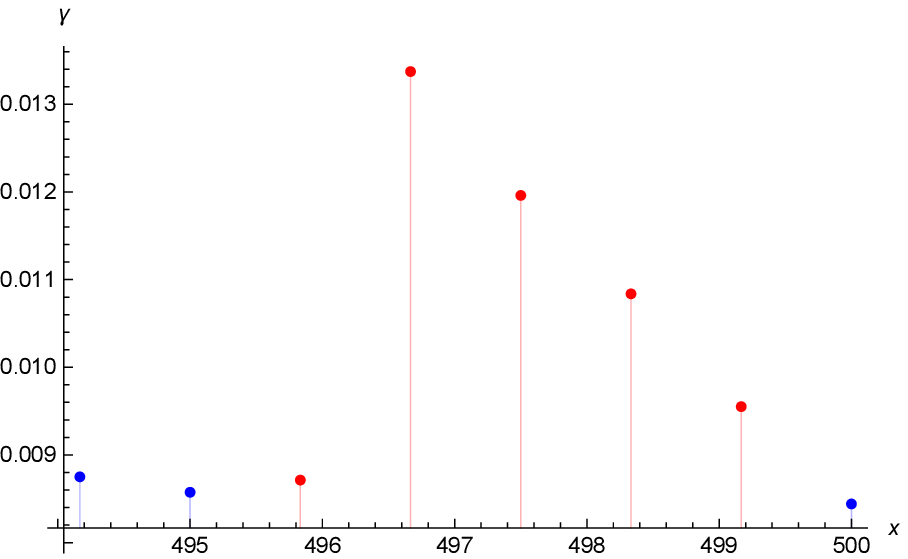}
		\end{minipage}
		\vspace{4ex}
		\captionof{figure}{Parameters $\gamma_i/M$ of the probability measure.}
		\label{Zooms}\end{figure}

\noindent We remark that the parameters grows up entering the data region before they start decreasing. A quick look at the solution dynamic equation, shows as that the $\gamma_i$ represent the market rigidity : higher is $\gamma_i$ more stable is the call price at the value $x_i$, we can see that the origin represent the most stable price (no one expect the price to vary there) and the region $[496.5,497]$ present more confidence in the market. Moreover, the parameters are much smaller than the classical model, except the origin, proving that the classical model represent an isotropic market with uniform uncertainty distribution.

	\vskip 1cm
	
\section{Discussion}	
	
\noindent The results established in this work demonstrate the superiority of the measure based Black-Scholes model in term of market data fit, while the classical model underestimate the option price.\\

\noindent The measure based Black-scholes offer a new information on the market consisting in the measure $\mu$ which can be interpreted as a market confidence measure : by considering the probability version $\nu=\displaystyle\frac{\mu}{M}=\sum \frac{\gamma_i}{M} \, \delta_i $, the parameter $\gamma_i$ represent the option price rigidity at the stock price $x_i$, higher is $\gamma_i$ more stable is the option price.\\

\noindent One remarkable fact is the parameters for the estimated model are lower than the uniform case of the Black-Scholes model, reflecting more instability in the market than the classical theory predict.\\

\noindent The model based on the measure $\mu$ open a new field of research by quantifying the market confidence (in opposition with market incertitude) : $\gamma(x)$ reflect the market confidence at the stock price $x$, inducing more rigidity and a slow change in the option price. This is just a measure of incertitude in the market, which was considered to be unmeasurable.\\

	\bibliographystyle{alpha}
	\bibliography{BibliographieClaire}
	
\end{document}